\documentclass[symmetry,article,accept,moreauthors,pdftex,10pt,a4paper]{mdpi} 
\pdfoutput=1
\SetLabelAlign{CenterWithParen}{\hfil({#1})\hfil}  
\usepackage{tabularx}
\usepackage{rotating}
\usepackage{amsmath,amssymb}
\usepackage[font=small,skip=0pt]{caption}

\firstpage{1} 
\pubvolume{10}
\issuenum{1}
\articlenumber{520}
\pubyear{2018}
\copyrightyear{2018}
\history{Received: 15 September 2018; Accepted: 15 October 2018; Published: 17 October 2018}

\Title{Dark Matter as a Non-Relativistic Bose--Einstein Condensate with Massive Gravitons}

\Author{{Emma Kun} $^{1}$\orcidA{}, Zolt\'{a}n Keresztes $^{1,}$*\orcidB{}, Saurya Das $^{2}$\orcidC{} and L\'aszl\'o \'A. Gergely$^{ 1}$\orcidD{}}

\AuthorNames{Emma Kun, Zolt\'{a}n Keresztes, Das Saurya, L\'aszl\'o \'A. Gergely}

\address{%
$^{1}$ \quad Institute of Physics, University of Szeged, D\'om t\'er 9,  Szeged H-6720, Hungary; kun@titan.physx.u-szeged.hu (E.K.); gergely@physx.u-szeged.hu (L.Á.G.)\\
$^{2}$ \quad Theoretical Physics Group and Quantum Alberta, Department of Physics and Astronomy, University of Lethbridge, 4401 University Drive, Lethbridge, AB T1K 3M4, Canada; saurya.das@uleth.ca \\
}

\corres{Correspondence: zkeresztes@titan.physx.u-szeged.hu}

\abstract{We confront a non-relativistic Bose--Einstein Condensate (BEC) model of light bosons interacting gravitationally either through a Newtonian or a Yukawa potential with the observed rotational curves of $12$ dwarf galaxies. The baryonic component is modelled as an axisymmetric exponential disk and its characteristics are derived from the surface luminosity profile of the galaxies. The purely baryonic fit is unsatisfactory, hence a dark matter component is clearly needed. The rotational curves of five galaxies could be explained with high confidence level by the BEC model. For these galaxies, we derive: (i) upper limits for the allowed graviton mass; and (ii) constraints on a velocity-type and a density-type quantity characterizing the BEC, both being expressed in terms of the BEC particle mass, scattering length and chemical potential. The upper limit for the graviton mass is of the order of
 $10^{-26}$  $\text{eV/c}^2$, three orders of magnitude stronger than the limit derived from recent gravitational wave detections.}


\begin{document}


\section{Introduction}
The universe is homogeneous and isotropic at scales greater than 
about $300$ Mpc.  
It is also spatially flat and expanding at an accelerating rate, following the laws of general relativity. 
The spatial flatness and 
accelerated expansion are most easily  
explained by assuming that the universe is almost entirely filled with just three constituents, namely visible matter, Dark Matter (DM) and Dark Energy (DE), with densities $\rho_{vis}$, $\rho_{DM}$ and $\rho_{DE}$, respectively, such that 
$ \rho_{vis} + \rho_{DM} + \rho_{DE}=
\rho_{crit}\equiv 3H_0^2/8\pi G \approx 10^{-26}~kg/m^3$
(where $H_0$ is the current value of 
the Hubble parameter and $G$ the
Newton's constant),
the so-called critical density, and 
$\rho_{vis}/\rho_{crit}=0.05,
\rho_{DM}/\rho_{crit}=0.25$ and
$\rho_{DM}/\rho_{crit}=0.70$~\cite{perlmutter,riess}.
It is the large amount of DE which causes the accelerated expansion. 
In other words, $95\%$ of its constituents
is invisible. 
Furthermore, the true nature of 
DM and DE remains to be understood.
There has been a number of promising candidates for DM, 
including weakly interacting massive particles (WIMPs), sterile neutrinos,
solitons, massive compact (halo) objects, 
primordial black holes, gravitons, etc., 
but none of them have been detected by dedicated experiments and some of them fail to accurately reproduce the rotation curves near galaxy centers
\cite{dmreview1,dmreview2}. 
Similarly, there has been a number of 
promising DE candidates as well, the most popular being a small cosmological constant, but any computation of the 
vacuum energy of quantum fields as a source of this constant gives incredibly large (and incorrect) estimates; another popular candidate is a dynamical scalar field \cite{dereview1,dereview2}. Two scalar fields are also able to model both DM and DE \cite{gergely2014}. Extra-dimensional modifications through a variable brane tension and five-dimensional Weyl curvature could also simulate the effects of DM and DE \cite{gergely2008}. In other theories, dark energy is the thermodynamic energy of the internal motions of a polytropic DM fluid~\cite{kleidis2015,kleidis2016}. Therefore, what exactly are DM and DE remain as two of the most important open questions in theoretical physics and cosmology. 

Given that DM pervades all universe, has mass and energy, gravitates and is cold (as otherwise it would not clump near galaxy centers), it was examined recently whether a Bose--Einstein condensate (BEC) of gravitons, axions or a Higgs type scalar can account for the DM content of our universe~\cite{db1,db2}. 
While this proposal is not new, and in fact BEC and superfluids as DM have been considered by various authors
\cite{Hu2000,Lopez,Bohua,
sudarshan1,sudarshan2,morikawa1,morikawa2,moffat,wang,boehmer1,boehmer2,sikivie,dvali,chavanis,kain,suarez,ebadi,laszlo1,bettoni,gielen,schive,davidson}, the novelty of the new proposal was twofold: 
(i) for the first time, it computed the quantum potential associated with the BEC; and (ii) it showed that this potential can in principle account for the DE content of our universe as well. 
It was also argued in the above papers that, if the BEC is accounting for DE
gravitons, then their mass would be tightly restricted to about $10^{-32}$~ $\text{eV/c}^2$. Any higher, and the corresponding Yukawa potential would be such that gravity would be shorter ranged than the current Hubble radius, about $10^{26}$ m, thereby contradicting cosmological observations.
Any lower and unitarity in a quantum field theory with gravitons would be lost \cite{dasaligravmass}. 

In this paper, we discuss the possibility of a BEC formed by scalar particles, interacting gravitationally through either the Newton or Yukawa potential. Such a BEC, interacting only through massless gravitons has been previously tested as a viable DM candidate by confronting with galactic rotation curves \cite{laszlo1,laszlo2}.

In this paper, we solve the time-dependent Scr\"odinger equation for the macroscopic wavefunction of a spherically symmetric BEC, 
where in place of the potential we plug-in a sum of the 
external gravitational potential and local density of the condensate, proportional to the absolute square of the wavefunction itself, 
times the self-interaction strength. The resultant non-linear Schr\"odinger equation is known as the 
Gross--Pitaevskii equation. 
For the self-interaction, we assume a two-body $\delta$-function type interaction
(the Thomas--Fermi approximation), 
while we assume that the external potential being massive-gravitational in nature, satisfying the Poisson equation with a mass term.
The BEC-forming bosons could be ultra-light, raising the question of why we use the non-relativistic Schr\"odinger equation. This is because, once in the condensate, they are in their ground states with little or no velocity, and hence non-relativistic for all practical purposes. 
Solving these coupled set of equations, we obtain the density function, the potential outside the condensate and   also the velocity profiles of the rotational curves. We then compare these analytical results with observational curves for $12$ dwarf galaxies and 
show that they agree with a high degree of confidence for five of them. For the remaining galaxies, no definitive conclusion can be drawn with a high confidence level. 
Nevertheless, our work provides the necessary 
groundwork and motivation to study the problem further to provide strong evidence for or against our model.

This paper is organized as follows. In the next section, we set the stage by summarizing the coupled differential equations that govern the BEC wavefunction and gravitational potential and find the BEC density profiles. 
In Section \ref{Section3}, we construct the corresponding analytical rotation curves.
In Section \ref{Section4}, we compare these and the rotational curves due to baryonic matter 
with the observational curves for galaxies. 
In Section \ref{Section5}, we find most probable bounds on the
graviton mass, as well as derive limits for a velocity-type and a density-type quantity characterizing the BEC.
 
\section{Self-Gravitating, Spherically Symmetric Bec Distribution in the Thomas-Fermi Approximation \label{BEC}}

A non-relativistic Bose--Einstein condensate in the mean-field approximation
is characterized by the wave function $\psi (\mathbf{r},t)$ obeying 
\begin{equation}
i\hbar \frac{\partial }{\partial t}\psi (\mathbf{r},t)=\left[ -\frac{\hbar
^{2}}{2m}\Delta +mV_{ext}\left( \mathbf{r}\right) +\lambda \rho \left( 
\mathbf{r},t\right) \right] \psi (\mathbf{r},t)~,  \label{Heisenberg}
\end{equation}%
known as the Gross--Pitaevskii equation \cite{Gross1,Gross2,Pitaevskii}. Here, $\hbar $ is the reduced Planck constant, $\mathbf{r}$ is
the position vector; $t$ is the time; $\Delta $ is the Laplacian; $m$ is the
boson mass;%
\begin{equation}
\rho \left( \mathbf{r},t\right) =\left\vert \psi (\mathbf{r},t)\right\vert
^{2}
\end{equation}%
is the probability density; the parameter $\lambda >0$ measures the atomic interactions and is also related to the scattering length \cite{rogersalazar2013}, characterizing the two-body interatomic potential energy:%
\begin{equation}
V_{self}=\lambda \delta \left( \mathbf{r}-\mathbf{r}^{\prime }\right);
\end{equation}%
and finally $V_{ext}\left( \mathbf{r}\right) $ is an external potential. For
a stationary state,
\begin{equation}
\psi (\mathbf{r},t)=\sqrt{\rho \left( \mathbf{r}\right) }\exp \left( \frac{%
i\mu }{\hbar }t\right) ~
\end{equation}%
where $\mu$ is a chemical potential energy \cite{rogersalazar2013,Giorgini1997}. When $\mu$ is constant, Equation 
(\ref{Heisenberg}) reduces to present
works \cite{boehmer1}, \cite{laszlo1}
\begin{equation}
mV_{ext}+V_{Q}+\lambda \rho =\mu ~,  \label{fieldEq}
\end{equation}%
where $V_{Q}$ is the quantum correction potential energy:%
\begin{equation}
V_{Q}=-\frac{\hbar ^{2}}{2m}\frac{\Delta \sqrt{\rho }}{\sqrt{\rho }}~.
\label{VQ}
\end{equation}%

We mention that  Equation (\ref{fieldEq}) is valid in the domain where $\rho \left( 
\mathbf{r}\right) \neq 0$.

The quantum correction $V_{Q}$ has significant contribution only
close to the BEC boundary \cite{wang}, therefore it can be neglected in comparison to
the self-interaction term $\lambda \rho $. This \textit{Thomas--Fermi
approximation} becomes increasingly accurate with an increasing number of
particles \cite{Liebetal}.

We assume $V_{ext}\left( \mathbf{r}\right) $ to be the gravitational potential
created by the condensate. In the case of massive gravitons, it
is described by the Yukawa-potential in the non-relativistic limit:%
\begin{equation}
V_{ext}=U_{Y}\left( \mathbf{r}\right) =-\int \frac{G\rho _{BEC}\left( 
\mathbf{r}^{\prime }\right) }{\left\vert \mathbf{r}-\mathbf{r}^{\prime
}\right\vert }e^{-\frac{\left\vert \mathbf{r}-\mathbf{r}^{\prime
}\right\vert }{R_{g}}}d^{3}\mathbf{r}^{\prime }\text{ },  \label{pot}
\end{equation}%
with $\rho _{BEC}=m\rho $, gravitational constant $G$, and characteristic
range of the force $R_{g}$ carried by the gravitons with mass $m_{g}$. The
relation between $R_{g}$ and $m_{g}$ is $R_{g}=\hbar /\left( m_{g}c\right) 
$, where $c$ is the speed of light and $\hbar $ is the reduced Planck
constant. The Yukawa potential obeys the following equation:%
\begin{equation}
\Delta U_{Y}-\frac{U_{Y}}{R_{g}^{2}}=4\pi G\rho _{BEC}~.  \label{YukawaDiff}
\end{equation}%
Contrary to Equation (\ref{fieldEq}),  Equation  (\ref{YukawaDiff}) is also valid in the
domain where $\rho \left( \mathbf{r}\right) =0$. In the massless graviton limit, we recover Newtonian gravity, in particular Equations  (\ref{pot}) and (\ref{YukawaDiff}) reduce to the Newtonian potential and Poisson equation. 

\subsection{Mass Density and the Gravitational Potential inside the Condensate%
}

The Laplacian of Equation (\ref{fieldEq})   using Equation  (\ref{YukawaDiff}) gives%
\begin{equation}
\Delta \rho _{BEC}+\frac{4\pi Gm^{2}}{\lambda }\rho _{BEC}=-\frac{m^{2}}{%
\lambda R_{g}^{2}}U_{Y}~.  \label{EndensEq}
\end{equation}%

For a spherical symmetric matter distribution, Equations (\ref{YukawaDiff}) and (%
\ref{EndensEq}) become%
\begin{equation}
\frac{d^{2}\left( rU_{Y}\right) }{dr^{2}}-\frac{1}{R_{g}^{2}}\left(
rU_{Y}\right) =4\pi G\left( r\rho _{BEC}\right) ~,  \label{diffeqrUy}
\end{equation}%
\begin{equation}
\frac{d^{2}\left( r\rho _{BEC}\right) }{dr^{2}}+\frac{1}{R_{\ast }^{2}}%
\left( r\rho _{BEC}\right) =-\frac{m^{2}}{\lambda R_{g}^{2}}\left(
rU_{Y}\right) ~.  \label{secondorder}
\end{equation}%
where we introduced the notation%
\begin{equation}
\frac{1}{R_{\ast }^{2}}=\frac{4\pi Gm^{2}}{\lambda }~.
\end{equation}%

This system gives the following fourth order, homogeneous, linear
differential equation for $r\rho _{BEC}$:%
\begin{equation}
\frac{d^{4}\left( r\rho _{BEC}\right) }{dr^{4}}+\Lambda ^{2}\frac{%
d^{2}\left( r\rho _{BEC}\right) }{dr^{2}}=0~,  \label{fourthorder}
\end{equation}%
with%
\begin{equation}
\Lambda =\sqrt{\frac{1}{R_{\ast }^{2}}-\frac{1}{R_{g}^{2}}}~.
\end{equation}

In the case of massless gravitons, $\pi R_{\ast }$ gives the radius of the
BEC halo \cite{laszlo1}. To have a real $\Lambda$, $%
R_{g}>R_{\ast }$ should hold, constraining the graviton mass from above. Typical dark matter halos have $\pi R_{\ast }$ of the order of $1$ kpc which
gives the following upper bound for the graviton mass: $m_{g}c^{2}<4\times 10^{-26}$~eV. Then, the
general solution of Equation (\ref{fourthorder}) is%
\begin{equation}
r\rho _{BEC}=A_{1}\sin \left( \Lambda r\right) +B_{1}\cos \left( \Lambda
r\right) +C_{1}r+D_{1}~.  \label{sol0}
\end{equation}%
with integration constants $A_{1}$, $B_{1}$, $C_{1}$ and $D_{1}$. This is why we impose the reality of $\Lambda$. For the imaginary case the general solution would contain runaway hyperbolic functions. This is 
also the solution of the system in \mbox{ Equations (\ref{diffeqrUy})--(\ref{secondorder}).}
Requiring $\rho _{BEC}$ to be bounded, we have $D_1=-B_1$. Then, the core density
of the condensate is%
\begin{equation}
0<\rho ^{\left( c\right) }\equiv \rho _{BEC}\left( r=0\right) =A_{1}\Lambda
+C_{1}~,
\end{equation}%
and the solution can be written as%
\begin{equation}
\rho _{BEC}\left( r\right) =\left( \rho ^{\left( c\right) }-C_{1}\right) 
\frac{\sin \left( \Lambda r\right) }{\Lambda r}+B_{1}\frac{\cos \left(
\Lambda r\right) -1}{r}+C_{1}~.  \label{sol}
\end{equation}

Substituting $\rho _{BEC}\left( r\right) $ in Equation (\ref{secondorder}), the
gravitational potential is%
\begin{equation}
-\frac{m^{2}}{\lambda R_{g}^{2}}\left( rU_{Y}\right) =\left( \rho ^{\left(
c\right) }-C_{1}\right) \frac{\sin \left( \Lambda r\right) }{\Lambda
R_{g}^{2}}+\frac{B_{1}}{R_{g}^{2}}\cos \left( \Lambda r\right) -\frac{B_{1}}{%
R_{\ast }^{2}}+\frac{C_{1}}{R_{\ast }^{2}}r~.  \label{potin}
\end{equation}%

Being related to the mass density by Equation (\ref{fieldEq}) gives%
\begin{equation}
B_{1}=0\text{~},~C_{1}=-\frac{m\mu }{\lambda R_{g}^{2}\Lambda ^{2}}~.
\label{BC}
\end{equation}

The BEC mass distribution ends at some radial distance $R_{BEC}$ (above which we set $\rho_{BEC}$ to zero), allowing to express $C_1$ in terms of $\rho^{\left( c\right) }$, $R_{BEC}$ and $\Lambda $ as%
\begin{equation}
C_{1}=\rho ^{\left( c\right) }\frac{\sin \left( \Lambda R_{BEC}\right) }{%
\Lambda R_{BEC}}\left( \frac{\sin \left( \Lambda R_{BEC}\right) }{\Lambda
R_{BEC}}-1\right) ^{-1}~.  \label{alpha}
\end{equation}

Finally, we consider the massless graviton limiting case $m_{g}\rightarrow 0$%
. Then, $R_{g}\rightarrow \infty $ implies $\Lambda =\sqrt{4\pi
Gm^{2}/\lambda }=1/R_{\ast }$ and $C_{1}=0$ (by Equation (\ref{BC})). Then, $\rho
_{BEC}\left( r\right) $ coincides with \mbox{{Equation (40)}~\cite{boehmer1}.}

\subsection{Gravitational Potential Outside the Condensate}

The potential $U$ is determined up to an arbitrary constant $A_2$, i.e.%
\begin{equation}
U^{out}=U_{Y}^{out}+A_{2}~.  \label{Uout}
\end{equation}%

Here, $U_{Y}^{out}$ satisfies Equation  (\ref{YukawaDiff}) with $\rho _{BEC}=0$. The solution for $U_{Y}^{out}$ is%
\begin{equation}
U_{Y}^{out}=B_{2}\frac{e^{-\frac{r}{R_{g}}}}{r}+C_{2}\frac{e^{\frac{r}{R_{g}}%
}}{r}~.  \label{UYout}
\end{equation}%

Since an exponentially growing gravitational potential is non-physical, $C_{2}=0$ and%
\begin{equation}
U^{out}=A_{2}+B_{2}\frac{e^{-\frac{r}{R_{g}}}}{r}~.  \label{Uout2}
\end{equation}

The constants $A_{2}$ and $B_{2}$ are determined from the junction
conditions: the potential is both continuous and continuously differentiable at $%
r=R_{BEC}$:%
\begin{eqnarray}
A_{2} &=&\frac{4\pi G\rho ^{\left( c\right) }}{1+\frac{R_{BEC}}{R_{g}}}\frac{%
R_{\ast }^{2}R_{g}^{2}}{1-\frac{\sin \left( \Lambda R_{BEC}\right) }{\Lambda
R_{BEC}}}\left[ \frac{\Lambda }{R_{g}}\sin \left( \Lambda R_{BEC}\right)
\right.  \nonumber \\
&&\left. \frac{1}{R_{\ast }^{2}}\frac{\sin \left( \Lambda R_{BEC}\right) }{%
\Lambda R_{BEC}}-\frac{\cos \left( \Lambda R_{BEC}\right) }{R_{g}^{2}}\right]
~,  \label{A2}
\end{eqnarray}%
\begin{equation}
B_{2}=\frac{4\pi G\rho ^{\left( c\right) }}{\frac{1}{R_{BEC}}+\frac{1}{R_{g}}%
}\frac{R_{\ast }^{2}}{1-\frac{\sin \left( \Lambda R_{BEC}\right) }{\Lambda
R_{BEC}}}\left[ \cos \left( \Lambda R_{BEC}\right) -\frac{\sin \left(
\Lambda R_{BEC}\right) }{\Lambda R_{BEC}}\right] e^{\frac{R_{BEC}}{R_{g}}}~.
\label{B2}
\end{equation}%

In the next section, we   see that the continuous differentiability of the
gravitational potential coincides with the continuity of the rotation curves.


\section{Rotation Curves in Case of Massive Gravitons} \label{Section3}

Newton's equation of motions give the velocity squared of stars in
circular orbit in the plane of the galaxy as%
\begin{equation}
v^{2}\left( R\right) =R\frac{\partial U}{\partial R}~.  \label{vcirc}
\end{equation}%

Here, $R$ is the radial coordinate in the galaxy's plane and $U$ is the
gravitational potential. In the case of massive gravitons, $U$ is given
by $U=U_{Y}+A$, where $U_{Y}$ satisfies the Yukawa-equation with the
relevant mass density and $A$ is a constant.

The contribution of the condensate to the circular velocity is%
\begin{equation}
v_{BEC}^{2}\left( R\right) =\frac{4\pi G\rho ^{\left( c\right) }R_{\ast }^{2}%
}{1-\frac{\sin \left( \Lambda R_{BEC}\right) }{\Lambda R_{BEC}}}\left[ \frac{%
\sin \left( \Lambda R\right) }{\Lambda R}-\cos \left( \Lambda R\right) %
\right]  \label{vBECsqin}
\end{equation}%
for $r\leq R_{BEC}$ and%
\begin{equation}
v_{BEC}^{2}\left( R\right) =-B_{2}\left( \frac{1}{R}+\frac{1}{R_{g}}\right)
e^{-\frac{R}{R_{g}}}  \label{vBECout}
\end{equation}%
for $r\geq R_{BEC}$.

In the relevant situations, the stars orbit inside the halo and their
rotation curves are determined by the parameters: $\rho ^{\left( c\right)
}R_{\ast }^{2}$, $R_{BEC}$ and $\Lambda $. In the limit $m_g \rightarrow 0$, the $v^2$ of the BEC with massless gravitons is recovered, given as {B{\"o}hmer} proposed  
 \cite{boehmer1}
\begin{equation}
v_{BEC}^2(R)=4\pi G \rho^{(c)} R_{\ast}^2 \left[ \frac{\sin (R_{\ast}^{-1} R)}{R_{\ast}^{-1} R} - \cos (R_{\ast}^{-1} R)\right]\label{eq:masslessbec}
\end{equation}
for $r\leq R_{BEC}$ and%
\begin{equation}
v_{BEC}^{2}\left( R\right) = 4G\rho^{(c)} \frac{R_{\ast}}{R}  \label{vBECout2}
\end{equation}%
for $r\geq R_{BEC}$.


\section{Best-Fit Rotational Curves} \label{Section4}

\subsection{Contribution of the Baryonic Matter in Newtonian and in Yukawa Gravitation}
\label{sectionbar}

The baryonic rotational curves are derived from the distribution of the luminous matter, given by the surface brightness $S = F /\Delta \Omega$ (radiative flux $F$ per solid angle
$\Delta \Omega$ measured in radian squared of the image) of the galaxy. 
The observed $S$ depends on the redshift as $1/(1+z)^4$, on the orientation of the galaxy rotational axis with respect to the line of sight of the observer, but independent from the curvature index of Friedmann universe. Since we investigate dwarf galaxies at small redshift ($z<0.002$), the $z$-dependence of $S$ is negligible. Instead of $S$ given in units of solar luminosity $L_\odot$ per square kiloparsec ($L_\odot/kpc^2$), the quantity $\mu$ given in units of $mag/arcsec^2$ can be employed, defined~through
\begin{equation}
S(R)=4.255\times 10^{14}\times 10^{0.4(\mathcal{M}_{\odot}-\mu(R))},
\label{eq:mutrafo}
\end{equation}
where $R$ is the distance measured the center of the galaxy in the galaxy plane and $\mathcal{M}_{\odot}$ is the absolute brightness of the Sun in units of $mag$. The absolute magnitude gives the luminosity of an object, on a logarithmic scale. It is defined to be equal to the apparent magnitude appearing from a distance of 10 parsecs. The bolometric absolute magnitude of a celestial object $\mathcal{M}_\star$, which takes into account the electromagnetic radiation on all wavelengths, is defined as $\mathcal{M}_\star-\mathcal{M}_\odot=-2.5 \log (L_\star/L_\odot)$, where $L_\star$ and $L_\odot$ are the luminosity of the object and of the Sun, respectively.

The brightness profile of the galaxies $\mu(R)$ was derived in some works
 \cite{Swaters1999,Swaters2002,Swaters2009} from isophotal fits, employing the orientation parameters of the galaxies (center, inclination angle and ellipticity). This analysis leads to $\mu(R)$ which would be seen if the galaxy rotational axis was parallel to the line-of-sight. We used this $\mu(R)$ to generate $S(R)$.

The surface photometry of the dwarf galaxies are consistent with modeling their baryonic component as an axisymmetric exponential disk with surface brightness \cite{Freeman1970}:
\begin{equation}
S(R)=S_0 \exp[{-R/b}]\label{eq:surfexpdisk}
\end{equation}
where $b$ is the scale length of the exponential disk, and $S_0$ is the central surface brightness. To convert this to mass density profiles, we fitted the mass-to-light ratio ($\Upsilon=M/L$) of the galaxies.

In Newtonian gravity, the rotational velocity squared of an exponential disk emerges as Freeman proposed
\cite{Freeman1970}:
\begin{equation}
v^2(R)=\pi G S_0 \Upsilon b \left(\frac{R}{b}\right)^2 (I_0 K_0-I_1 K_1), \label{eq:newtonexpdisk}
\end{equation}
with $I$ and $K$ the modified Bessel functions, evaluated at $R/2b$. In Yukawa gravity, a more cumbersome expression has been given in the work of De Araujo and Miranda
\cite{deAraujo2007} as
\begin{eqnarray}
v^2(R)&=&2\pi G S_0 \Upsilon R \times \left[ \int^{\infty}_{b/\lambda} \frac{\sqrt{x^2-b^2/\lambda^2}}{(1+x^2)^{3/2}} J_1\left( \frac{R}{b}\sqrt{x^2-b^2/\lambda^2} \right)dx\right.  \nonumber \\
&&\left. -\int^{b/\lambda}_{0} \frac{\sqrt{b^2/\lambda^2-x^2}}{(1+x^2)^{3/2}} I_1\left( \frac{R}{b}\sqrt{b^2/\lambda^2-x^2} \right)dx  \right], \label{eq:yukexpdisk}
\end{eqnarray}
where $\lambda=h/m_g/c=2\pi R_g$ is the Compton wavelength. For $b/\lambda \ll 1$, the Newtonian limit is~recovered.

 \subsection{Testing Pure Baryonic and Baryonic + Dark Matter Models}

We chose 12 late-type dwarf galaxies from the Westerbork HI survey of spiral and irregular galaxies \citep{Swaters1999,Swaters2002,Swaters2009} to test rotation curve models. The selection criterion was that these disk-like galaxies have the longest $R$-band surface photometry profiles and rotation curves. For the absolute $R$-magnitude of the Sun, $\mathcal{M}_{\odot,R}=4.42^m$ \cite{Binney1998} was adopted. Then, we fitted Equation (\ref{eq:surfexpdisk}) to the surface luminosity profile of the galaxies, calculated with Equation (\ref{eq:mutrafo}) from $\mu(R)$. The best-fit parameters describing the photometric profile of the dwarf galaxies are given in Table \ref{table:vrot_bestfit1}.

We derived the pure baryonic rotational curves by fitting the square root of Equation (\ref{eq:newtonexpdisk}) to the observed rotational curves allowing for variable $M/L$. The pure baryonic model leads to best-fit model-rotation curves above $5\sigma$ significance level for all galaxies (the $\chi^2$-s are presented in the first group of columns in Table \ref{table:vrot_bestfit1}), hence a dark matter component is clearly required.

\begin{figure} [H]
\centering

\includegraphics[width=125pt,height=85pt]{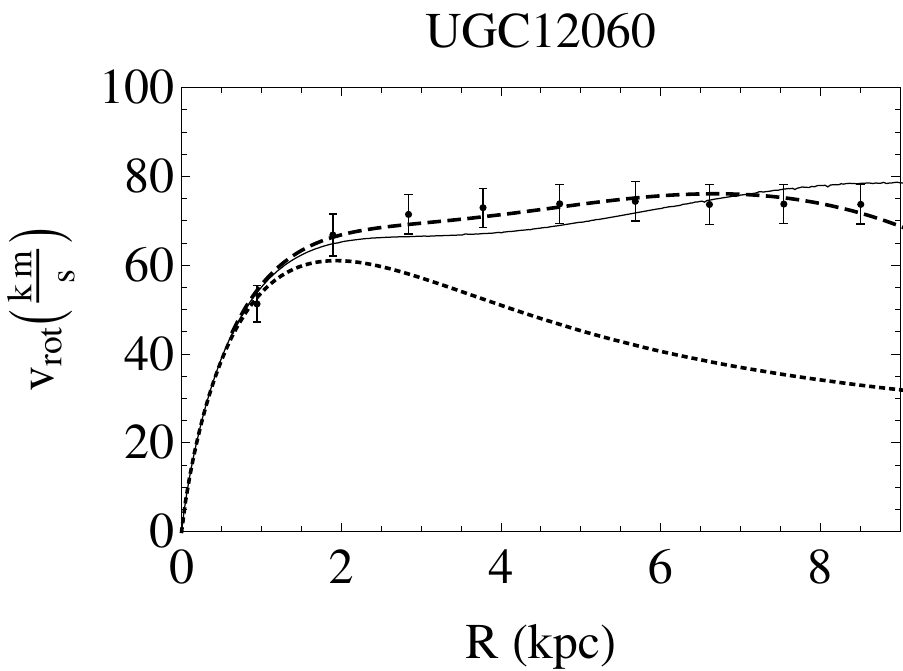}
\includegraphics[width=125pt,height=85pt]{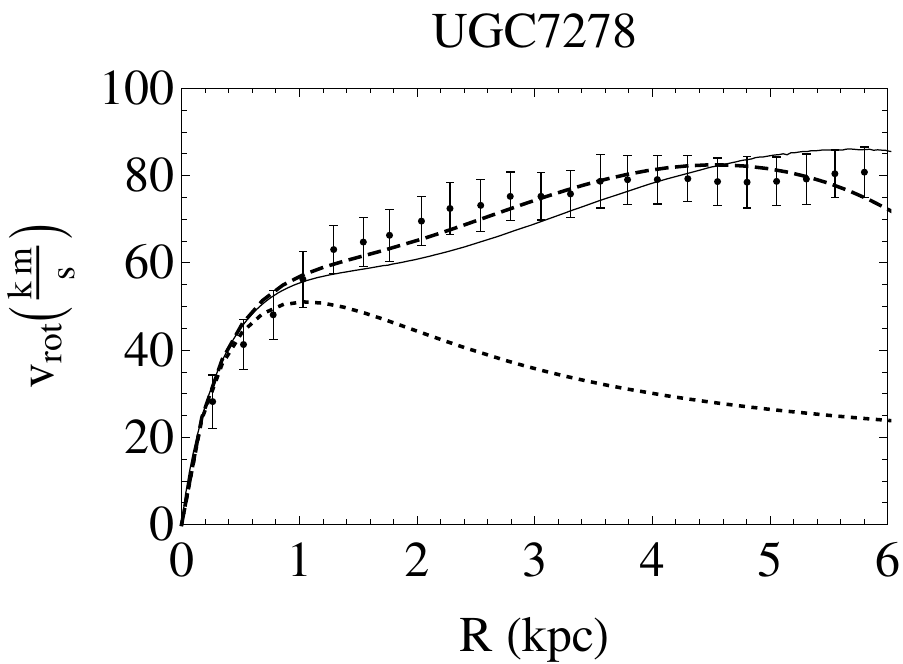}
\includegraphics[width=125pt,height=85pt]{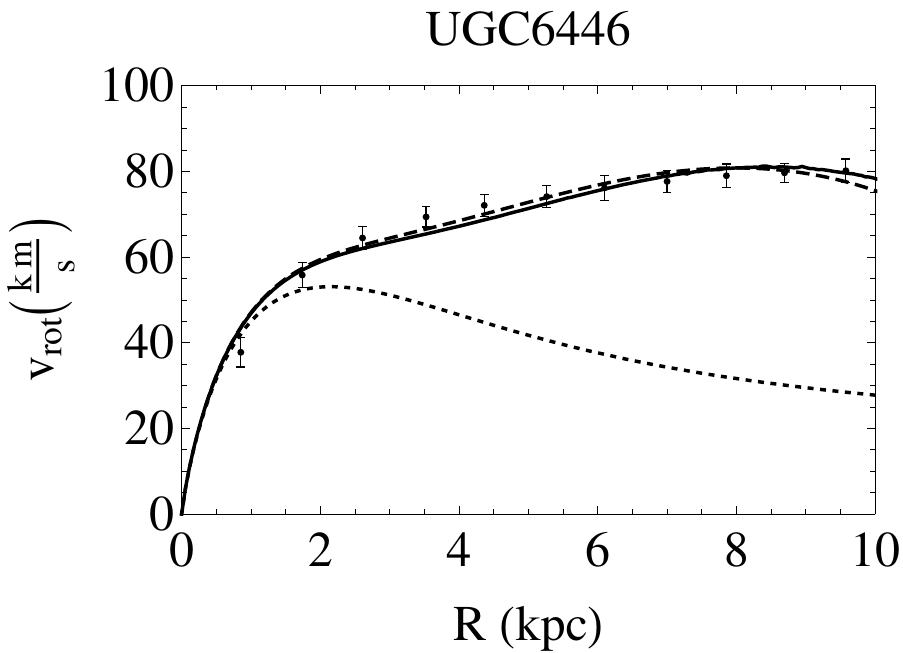}\newline
\includegraphics[width=125pt,height=85pt]{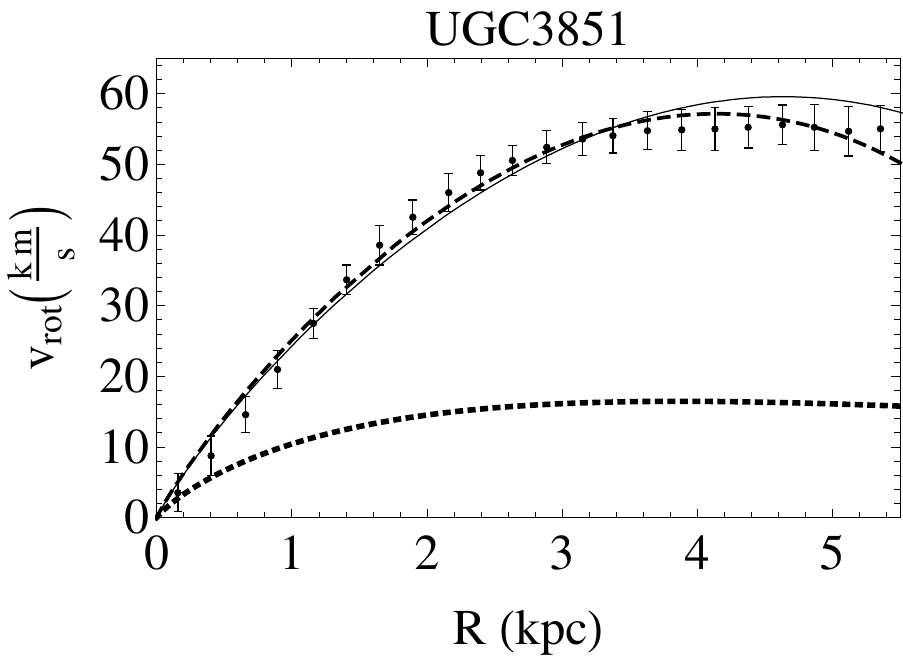}
\includegraphics[width=125pt,height=85pt]{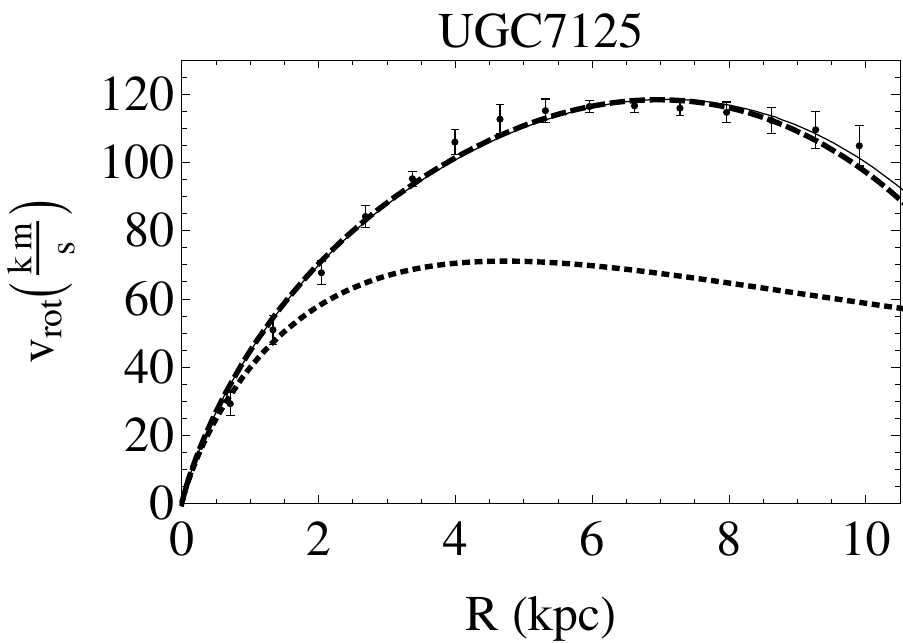}
\includegraphics[width=125pt,height=85pt]{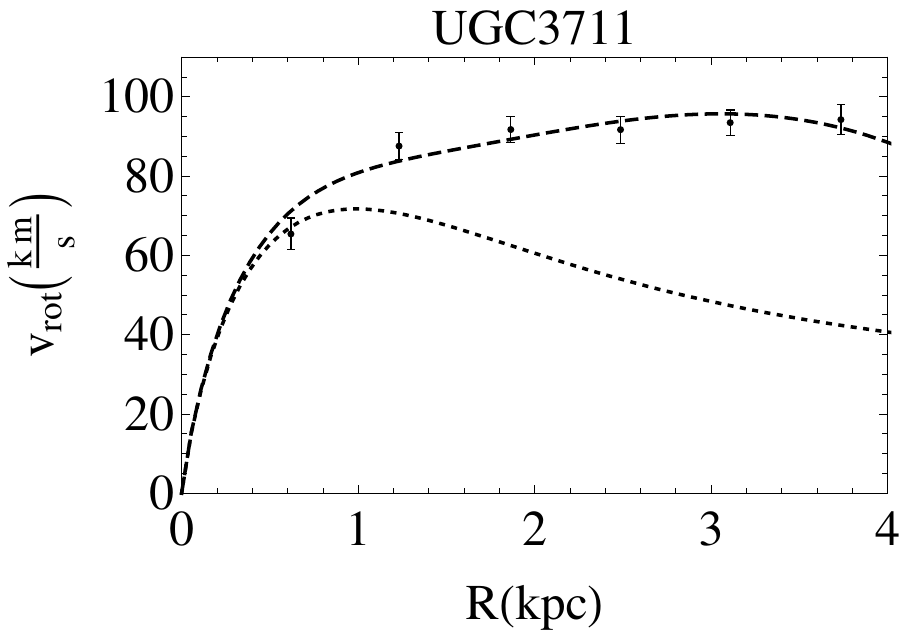}\newline
\includegraphics[width=125pt,height=85pt]{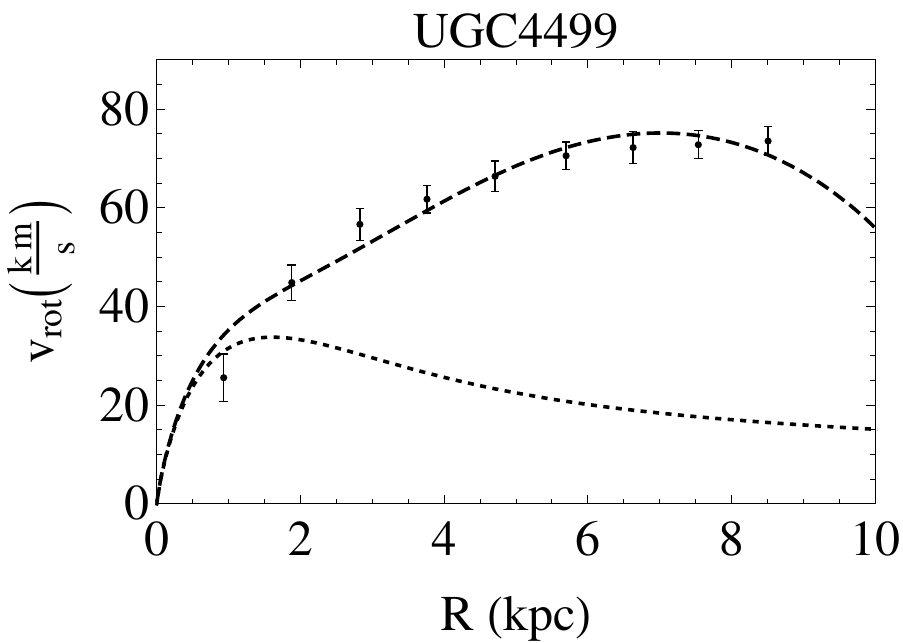}
\includegraphics[width=125pt,height=85pt]{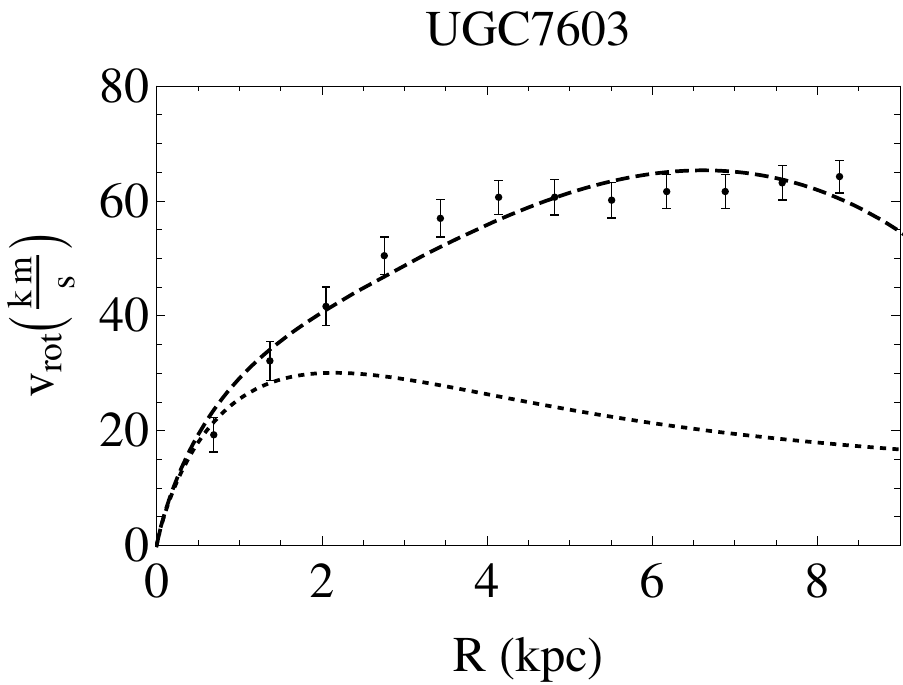}
\includegraphics[width=125pt,height=85pt]{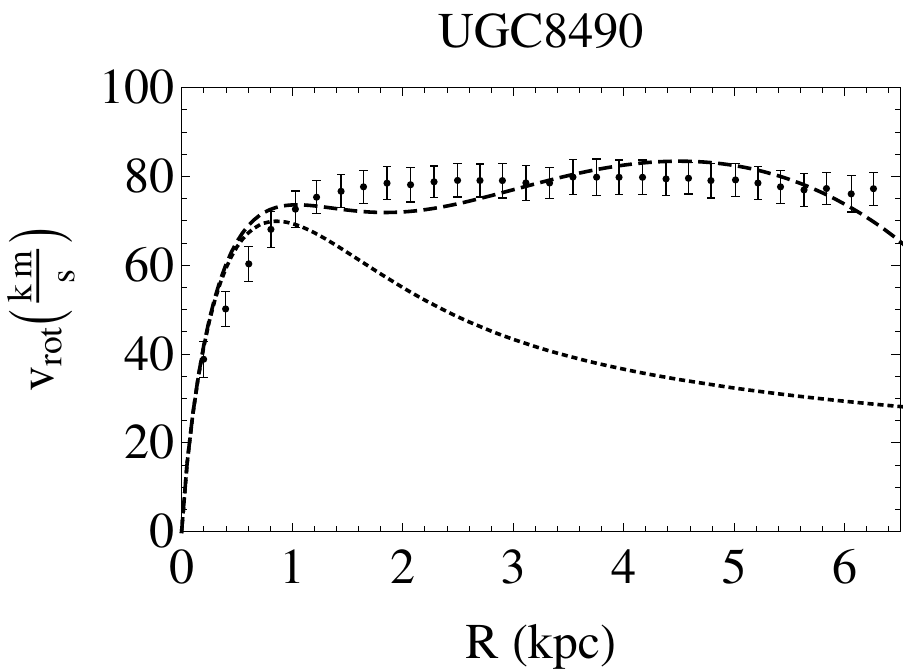}\newline
\includegraphics[width=125pt,height=85pt]{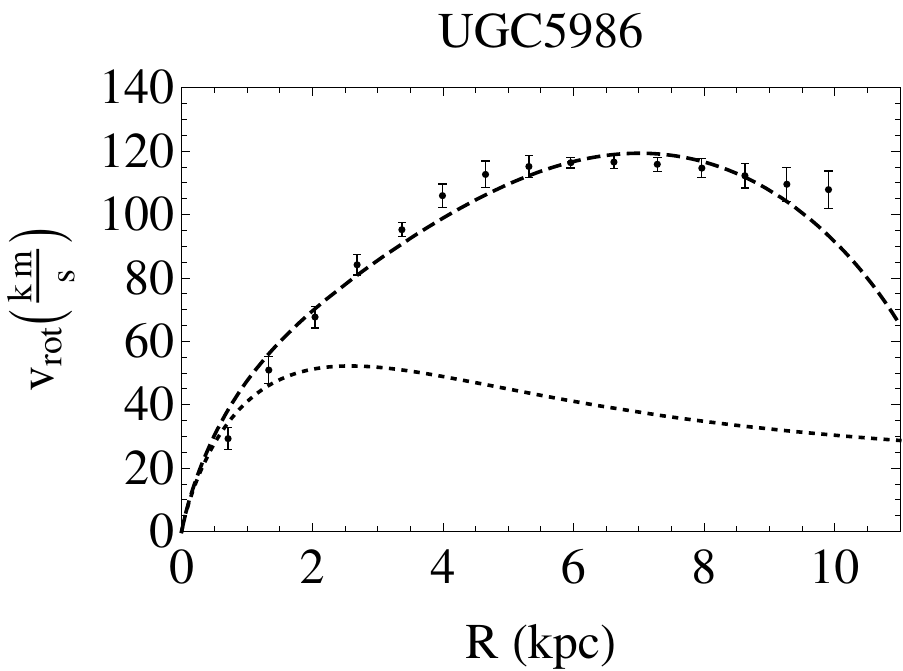}
\includegraphics[width=125pt,height=85pt]{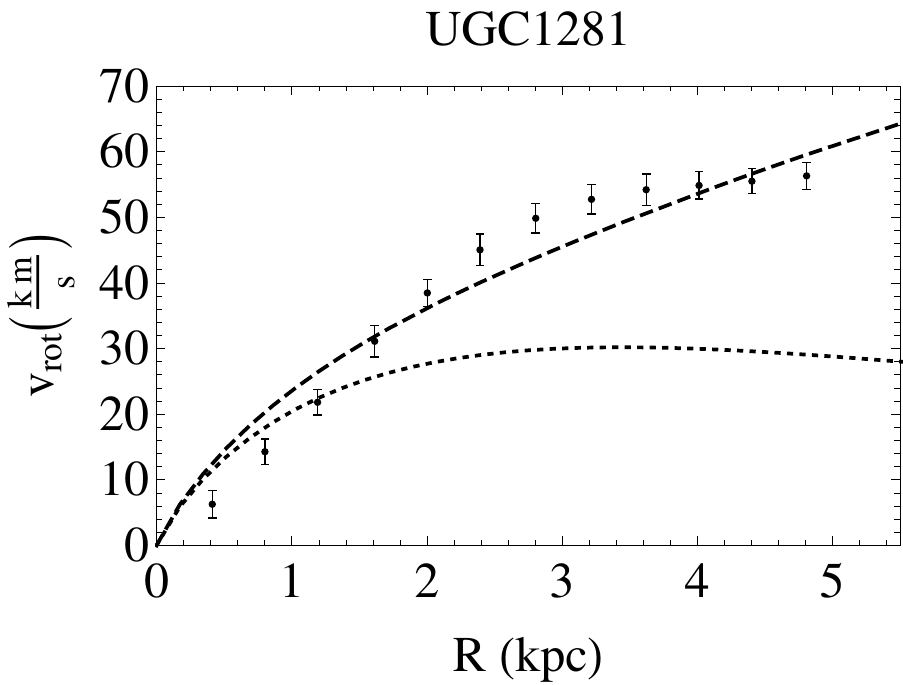}
\includegraphics[width=125pt,height=85pt]{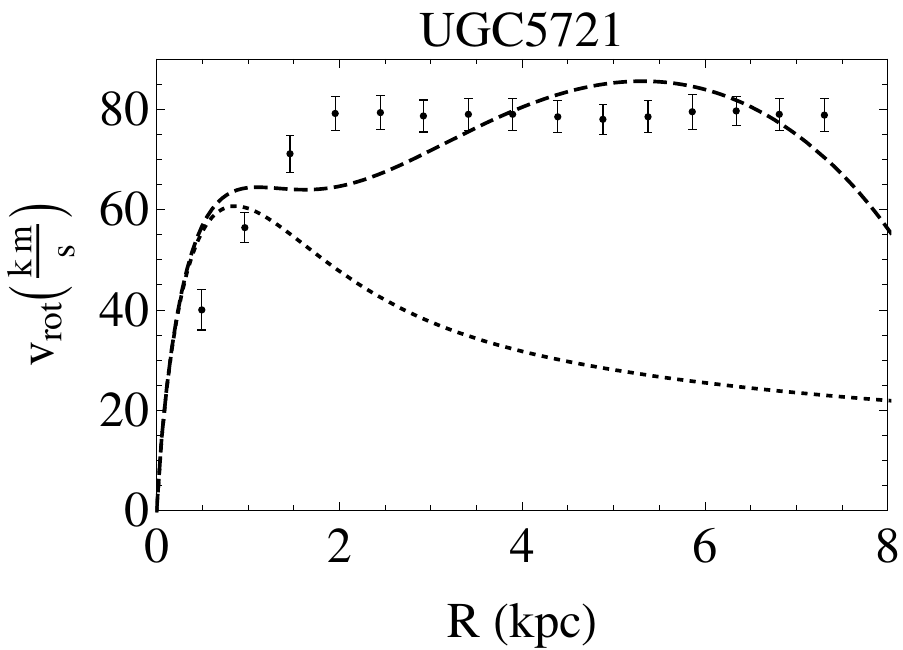}\newline
\caption{Theoretical rotational curves of the dwarf galaxy sample. The dots with error-bars denote archive rotational velocity curves. The model rotation curves are denoted as follows: pure baryonic in Newtonian gravitation with dotted line, baryonic + BEC with massless gravitons in Newtonian gravitation with dashed line, and baryonic + BEC with the upper limit on $m_g$ in Yukawa gravitation with continuous line.}
\vspace{12pt} 
\label{fig:vrot_dwarfs}
\end{figure}

Then, we fitted theoretical rotation curves with contributions of baryonic matter and BEC-type dark matter with massless gravitons to the observed rotational curves in Newtonian gravity. The model--rotational velocity of the galaxies in this case is given by the square root of the sum of velocity squares given by Equations (\ref{eq:masslessbec}) and (\ref{eq:newtonexpdisk}) with free parameters $\Upsilon$, $\rho^{(c)}$ and $R_\ast$. The best-fit parameters are given in the second group of columns of Table \ref{table:vrot_bestfit1}. Adding the contribution of a BEC-type dark matter component with zero-mass gravitons to rotational velocity significantly improves the $\chi^2$ for all galaxies, as well as results in smaller values of M/L. The fits are within $1\sigma$ significance level in five cases (UGC3851, UGC6446, UGC7125, UGC7278, and UGC12060), between $1\sigma$ and $2\sigma$ in three cases (UGC3711, UGC4499, and UGC7603), between $2\sigma$ and $3\sigma$ in one case (UGC8490), between $3\sigma$ and $4\sigma$ in one case (UGC5986) and above $5\sigma$ in two cases (UGC1281 and UGC5721). We note that the bumpy characteristic of the BEC model results in the limitation of the model in some cases, the decreasing branch of the theoretical rotation curve of the BEC component being unable to follow the observed plateau of the galaxies (UGC5721, UGC5986, and UGC8490). The theoretical rotation curves composed of a baryonic component plus BEC-type dark matter component with massless gravitons are presented on Figure \ref{fig:vrot_dwarfs}.

\newpage
\paperwidth=\pdfpageheight
\paperheight=\pdfpagewidth
\pdfpageheight=\paperheight
\pdfpagewidth=\paperwidth
\newgeometry{layoutwidth=297mm,layoutheight=210 mm, left=2.7cm,right=2.7cm,top=1.8cm,bottom=1.5cm, includehead,includefoot}
\fancyheadoffset[LO,RE]{0cm}
\fancyheadoffset[RO,LE]{0cm}

\begin{table}
\caption{Parameters describing the theoretical rotational curve models of the 12 dwarf galaxies. Best-fit parameters of the pure baryonic model in the first group of columns: central surface brightness $S_0$, scale parameter $b$, $M/L$ ratio $\Upsilon$, along with the $\chi^2$ of the fit. This model results in best-fit model-rotation curves above $5\sigma$ significance level for all galaxies. Best-fit parameters of the baryonic matter + BEC with massless gravitons appear in the second group of columns: $M/L$ ratio $\Upsilon$, characteristic density $\rho^{(c)}$, distance parameter $R_\ast$, along with the $\chi^2$ of the fit and the respective significance levels. Constraints on the parameter $m^2/\lambda$ are also derived. In five cases, the fits $\chi^2$ are within $1\sigma$ and marked as boldface. The fits are between $1\sigma$ and $2\sigma$ in three cases, between $2\sigma$ and $3\sigma$ in one case, between $3\sigma$ and $4\sigma$ in one case and above $5\sigma$ in two cases. Best-fit parameters of the baryonic matter + BEC with massive gravitons are given in the third group of columns only for the well-fitting galaxies: the range for $R_{BEC}$ and the upper limit on $m_g$ are those for which the fit remains within $1\sigma$. Corresponding constraints on the parameter $m/\mu$ are also derived.} 
\vspace{6pt} 
\label{table:vrot_bestfit1}
\centering
\begin{tabular}{ccccccccccccccc}
\toprule
& \multicolumn{4}{c}{\textbf{Pure Baryonic}} & \multicolumn{6}{c}{\textbf{Baryonic + BEC with} \boldmath{$m_g=0$}}& \multicolumn{3}{c}{\textbf{Baryonic + BEC with}\boldmath{ $m_g>0$}}\\
\boldmath{$ID$ }& $S_0$ & $b$ & $\Upsilon$ & $\chi^2$ & $\Upsilon$ & $\rho^{(c)}$ & $R_\ast$ &  $\frac{m^2}{\lambda}$ & $\chi^2$ & sign. lev. & $R_{BEC}$ & $m_g$ & $\frac{m}{\mu}$ & sign. lev.\\
 & $10^8 \frac{L_\odot}{kpc^2}$ & $kpc$ &   & & & $10^7 \frac{M_\odot}{kpc^3}$ & $kpc$ & $10^{-31} \frac{kg s^2}{m^5}$ & & & $kpc$ & $10^{-26} \frac{eV}{c^2}$  &  $10^{-10} \frac{s^2}{m^2}$ & \\
\midrule
 UGC12060& 0.7 & 0.90 & 11.23& 155 & 5.50 $\pm$ 0.33 & 1.07 $\pm$ 0.11 & 2.650 $\pm$ 0.118 & 1.78 $\pm$ 0.16 & \textbf{1.69} & $1\sigma=5.89$ & [7.3 $\div$ 10.6] & $<0.95$&$<7.02$ & $1\sigma=7.08$ \\
 UGC7278 & 6.1 & 0.49 & 2.59 & 499 & 0.81 $\pm$ 0.06 & 3.53 $\pm$ 0.23 & 1.702 $\pm$ 0.048 & 4.32 $\pm$ 0.24 & \textbf{7.91} & $1\sigma=21.36$ & [4.6 $\div$ 6.8] & $<1.40$  & $<5.46$ & $1\sigma=22.44$ \\
   UGC6446 & 1.9 & 1.00 & 3.89 & 809 & 1.37 $\pm$ 0.11 & 1.02 $\pm$ 0.09 & 3.040 $\pm$ 0.128 & 1.36 $\pm$ 0.11 & \textbf{7.91} & $1\sigma=8.18$ & [9.2 $\div$ 10] & $<0.42$  & $<4.27$ & $1\sigma=9.86$  \\
  UGC3851 & 0.5 & 1.80 & 2.74 & 86  & 0.74 $\pm$ 0.18 & 1.91 $\pm$ 0.22 & 1.509 $\pm$ 0.038 & 5.50 $\pm$ 0.28 & \textbf{11.30} & $1\sigma=20.28$ & [4.3 $\div$ 5.5]& $<1.26$  & $<11.4$ & $1\sigma=21.36$  \\
 UGC7125 & 1.2 & 2.20 & 4.50 & 285 & 1.78 $\pm$ 0.18 & 2.26 $\pm$ 0.21 & 2.670 $\pm$ 0.071 & 1.76 $\pm$ 0.93 & \textbf{11.82} & $1\sigma=12.64$ & [8.2 $\div$ 8.6] &  $<0.31$  & $<2.44$& $1\sigma=13.74$ \\ 
\midrule
 UGC3711 & 5.2 & 0.46 & 4.40 & 232 & 2.00 & 8.06 & 1.212 & - & 5.11 & $2\sigma=6.18$ & -& - & -  & -\\
 UGC4499 & 1.4 & 0.75 & 6.30 & 603 & 1.00 & 1.34 & 2.590 &- & 8.51 & $2\sigma=11.31$ & - & -  & - & - \\
 UGC7603 & 2.1 & 1.00 & 1.88 & 462 & 0.40 & 1.07 & 2.470 &- & 13.46 & $2\sigma=15.78$ & - & -  & - & - \\
\midrule
 UGC8490 & 2.8 & 0.40 & 9.52 & 1350& 4.06 & 3.35 & 1.715 & -& 40.27 & $3\sigma=50.55$ & - & -  & -& - \\
\midrule
UGC5986 & 4.4 & 1.20 & 3.95 & 1682& 0.48 & 3.17 & 2.620 &- & 32.12& $4\sigma=38.54$ & - & - & -& - \\
\midrule
UGC1281 & 1.0 & 1.60 & 1.33 & 231 & 0.53 & 0.75 & 3.70 &- & 48.74 & $5\sigma=43.98$ & - & -&- & -\\
 UGC5721 & 4.9 & 0.40 & 5.79 & 1388& 1.75 & 2.84 & 1.982 &- & 88.56& $5\sigma=50.21$ & - & - & -& - \\
\bottomrule
\end{tabular}

\end{table}
\newpage
\restoregeometry
\paperwidth=\pdfpageheight
\paperheight=\pdfpagewidth
\pdfpageheight=\paperheight
\pdfpagewidth=\paperwidth
\headwidth=\textwidth

We attempted to distinguish among galaxies to be included in well-fitting or less well-fitting classes based on their baryonic matter distribution. Several factors affect the goodness of the fits, as follows. The best-fit falls outside the $1\sigma$ significance level in the case of seven galaxies. Among these galaxies, UGC8490 and UGC5721 have ($a_1$) steeply rising rotational curve due to their centralized baryonic matter distribution ($b < 0.5$ kpc, $v_{max}>50 km s^{-1}$) with ($a_2$) long, approximately constant height observed plateau. Joint fulfilment of these criteria does not occur for the well-fitting galaxies, as $b\gtrsim0.5$ kpc for them. The rest of the galaxies with best-fits falling outside the $1\sigma$ significance level have ($b_1$) slowly rising rotational curve due to their less centralized baryonic matter distribution ($b>0.5$ kpc, $v_{max}<50 km s^{-1}$) with ($b_2$) short, variable height observed plateau, holding relatively small number of observational points ($N\leq 15$, a small $N$ lowers the $1\sigma$ significance level). The well-fitting galaxies do not belong to this group, as either they hold more observational points, or have a longer, approximately constant height observed plateau. We expect that for the galaxies not falling in the classes with baryonic and observational characteristics summarized by either properties ($a_1$)--($a_2$) or ($b_1$)--($b_2$) the BEC dark matter model represents a good fit. Finally, we note the galaxy UGC3711 represents a special case due to the lack of sufficient observational data. Although the shape of its rotational curve is very similar to that of the best-fitting galaxy, UGC12060, it is based on just six observational points, lowering the $1\sigma$ level. Its points also have smaller error bars, which increases the $\chi^2$. This results in the best-fit rotational curve of UGC3711 falling outside out the $1\sigma$ significance level. 

Finally, we fitted the theoretical rotational curves given by both a baryonic component and a non-relativistic BEC component with massive gravitons, employing Yukawa gravity. The parameters $\Upsilon$, $\rho^{(c)}$ and $R_\ast$ were kept from the best-fit galaxy models composed of baryonic matter + BEC with massless gravitons. The model--rotational velocity of the galaxies arises as the square root of the sum of velocity squares given by Equations (\ref{vBECsqin}) and (\ref{eq:yukexpdisk}) with free parameters $R_{BEC}$ and $R_g$. Adding mass to the gravitons in the BEC model leads to similar performances of the fits.


\section{Discussion and Concluding Remarks} \label{Section5}

We estimated the upper limit on the graviton mass, employing first the theoretical condition of the existence of the constant $\Lambda$, then analyzing the modelfit results of those five dwarf galaxies for which the fit of the BEC model with massive gravitons to data was within $1\sigma$ significance level. 

Keeping the best-fit parameters $\rho^{(c)}$, $R_\ast$, we varied the value of $R_{BEC}$ and $R_g$ and calculated the $\chi^2$ between model and data. The upper limit on the graviton mass $m_g$ has been estimated from the values of $R_g$, for which $\chi^2=1\sigma$ has been reached. The results are given in Table \ref{table:vrot_bestfit1}. We plotted the theoretical rotation curves given by a baryonic plus a non-relativistic BEC component with massive gravitons with limiting mass in Figure \ref{fig:vrot_dwarfs}. As shown in Table \ref{table:constraints}, the fit with the rotation curve data has improved the limit on the graviton mass in all cases.

\begin{table}[H]
\caption{Constraints for both the upper limit for the mass of the graviton (first from the existence of $\Lambda$, second from the rotation curves) and for the velocity-type and density-type BEC parameters (related to the mass of the BEC particle, scattering length and chemical potential) in the case of the five well-fitting galaxies.}
\centering
\label{table:constraints}
\begin{tabular}{ccccc}
\toprule
\boldmath{$ID$} & \boldmath{$m_g (\Lambda \in {\rm I\!R})$} & \boldmath{$m_g$ }& \boldmath{$\bar{v}_{BEC}$ }& \boldmath{$\bar{\rho}_{BEC}$} \\
& \boldmath{$10^{-26} \frac{eV}{c^2}$ }&\boldmath{ $10^{-26} \frac{eV}{c^2}$ }&\boldmath{ $\frac{m}{s} $} &\boldmath{ $10^6 \frac{M_\odot}{kpc^3}$} \\
\midrule
UGC12060 & $<1.51$ & $<0.95$ & $37724$ & $3.75$ \\
UGC7278 & $<2.35$ & $<1.40$ & $42800$ & $11.69$ \\
UGC6446 & $<1.32$ & $<0.42$ & $48383$ & $4.68$ \\
UGC3851 & $<2.65$ & $<1.26$ & $29571$ & $7.1$ \\
UGC7125 & $<1.5$ & $<0.31$ & $63964$ & $10.61$ \\
\bottomrule
\end{tabular}
\end{table}

Comparing the theoretical rotation curves derived in our model with the observational ones, we 
found the upper limit to the graviton mass to be of the order of $10^{-26}$~$\text{eV/c}^2$ . We also note that the constraint on the graviton mass imposed from the dispersion relations tested by the first three observations of gravitational waves, $7.7 \times 10^{-23}$~ $\text{eV/c}^2$ \cite{GW170104}, is still weaker than the present one.

For the BEC, we could derive two accompanying limits: (i) first $m^2/\lambda$ has been constrained from the corresponding values of $R_\ast$ arising from the fit with the massless gravity model; and then (ii) $m/\mu$ has been constrained from the constraints derived for the graviton mass and our previous fits through Equations (\ref{BC}) and (\ref{alpha}). 
These are related to the bosonic mass, chemical potential and scattering length, but only two combinations of them, a velocity-type quantity
\begin{equation}
\bar{v}_{BEC}=\sqrt{\frac{\mu}{m}}
\end{equation}
and a density-type quantity
\begin{equation}
\bar{\rho}_{BEC}=\frac{m^2}{\lambda} \bar{v}_{BEC}^2
\end{equation}
were restricted, both characterizing the BEC. Their values are also given in Table \ref{table:constraints} for the set of five well-fitting galaxies.

If the BEC consists of massive gravitons with the limiting masses $m=m_g$ determined in Table \ref{table:constraints}, the chemical potential $\mu$ and the constant characterizing the interparticle interaction $\lambda$ can be determined as presented in Table \ref{table:mulambdaconstraints}. 

\begin{table}[H]
\caption{Constraints on $\mu$ and $\lambda$ assuming $m=m_g$ in case of the five well fitting galaxies.}
\centering
\label{table:mulambdaconstraints}
\begin{tabular}{ccc}
\toprule
\boldmath{$ID$} & \boldmath{$\mu (m=m_g)$} & \boldmath{$\lambda (m=m_g)$ } \\
 & $10^{-53} \frac{m^2}{s^2} kg$ & $10^{-94} \frac{m^5}{s^2}kg$ \\
\midrule
UGC12060 & $<2.41$ & $<16.08$\\
UGC7278 & $<4.57$ & $<14.40$\\
UGC6446 & $<1.75$ & $<4.14$\\
UGC3851 & $<1.96$ & $<9.17$\\
UGC7125 & $<2.26$ & $<1.74$\\
\bottomrule
\end{tabular}
\end{table}

With this, we established observational constraints for both the upper limit for the mass of the graviton and for the BEC.


\vspace{6pt} 



\authorcontributions{Conceptualization, L.{\'A}.G. and S.D.; Data curation, E.K.; Formal analysis, L.{\'A}.G., E.K. and
Z.K.; Funding acquisition, L.{\'A}.G., Z.K. and S.D.; Investigation, E.K.; Methodology, L.{\'A}.G., E.K. and Z.K.; Software,
E.K.; Supervision, L.{\'A}.G. and Z.K.; Validation, L.{\'A}.G., Z.K. and S.D.; Visualization, E.K.; Writing—original draft,
L.{\'A}.G., E.K., Z.K. and S.D.;Writing—review \& editing, L.{\'A}.G.}

\funding{This work was supported by the Hungarian National Research Development and Innovation Office
(NKFIH) in the form of the grant 123996 and by the Natural Sciences and Engineering Research Council of Canada
and based upon work from the COST action CA15117 (CANTATA), supported by COST (European Cooperation
in Science and Technology). The work of Z.K. was also supported by the J{\'a}nos Bolyai Research Scholarship of the
Hungarian Academy of Sciences and by the UNKP-18-4 New National Excellence Program of the Ministry of
Human Capacities.}

\acknowledgments{This work was supported by the Hungarian National Research Development and Innovation
Office (NKFIH) in the form of the grant 123996 and by the Natural Sciences and Engineering Research Council
of Canada and based upon work from the COST action CA15117 (CANTATA), supported by COST (European
Cooperation in Science and Technology). The work of Z.K. was also supported by the J{\'a}nos Bolyai Research
Scholarship of the Hungarian Academy of Sciences and by the UNKP-18-4 New National Excellence Program of
the Ministry of Human Capacities.}

\conflictsofinterest{The founding sponsors had no role in the design of the study; in the collection, analyses, or interpretation of data; in the writing of the manuscript, or in the decision to publish the results.}
\reftitle{References}





\end{document}